# US Presidential Election 2012 Prediction using Census Corrected Twitter Model[1]

Murphy Choy
Michelle Cheong
Ma Nang Laik
Koo Ping Shung

**Abstract**

US Presidential Election 2012 has been a very tight race between the two key candidates. There were intense battle between the two key candidates. The election reflects the sentiment of the electorate towards the achievements of the incumbent President Obama. The campaign lasted several months and the effects can be felt in the internet and twitter. The presidential debates injected new vigor in the challenger's campaign and successfully captured the electorate of several states posing a threat to the incumbent's position. Much of the sentiment in the election has been captured in the online discussions. In this paper, we will be using the original model described in Choy et. al. (2011) using twitter data to forecast the next US president.

**Introduction**

Social media has been widely used by politicians to increase the reach of their campaigns. The campaign of President Obama in 2008 demonstrated the power and reach of social media and how it can be used to change the political landscape. Social media platforms have been touted as the leading platform to engage voters. Political analysts (Stirland, 2008; Pasek, 2006; Xenos, 2007) attributed the success to the active and effective use of social media platforms to engage voters notably the younger generations usually ignored or given less importance (Pasek, 2006) by veteran politicians. The use of social media also helped to reach small population such as the Hispanics and activate support for the candidates. The poor management of social media platforms and absymal out reaches by McCain and Palin improved Obama's overall advantage in attracting voters (Stirland, 2008) and encouraging their turnout. Twitter has become a popular tool of choice for politicians to spread influence.

Twitter was developed as a micro-blogging tool where all the status updates as well as the opinions written can be searched and extracted using the twitter search API (Twitter, 2011). Search capabilities allow almost real life time viewing of information posted online by various users. Political associations as well as various activitist groups have successfully used it to voice their opinions, political positions as well as gathering supports from the online audiences. Every campaign has an online presence and the research focus on the formulation of these successful online campaigns. Campaigners are on a constant lookout for new social media channels to garner supporters and drive voter shares. There are many analysts and skeptics who believe that twitter is not very useful (Pearanalytics, 2009; Gayo-Avello, 2012), others have lauded the immense potential of twitter (Skemp ,2009).

There are three major objectives to be achieved using twitter information. The first aim is to assess the validity of the framework proposed in Choy et. al. (2011) to predict the new President. The second aim is to assess the methodology's power to calculate and predict state level electorate results. Finally, we will also like to evaluate the impact of twitter on the greater internet as a whole.

**Background to the US Presidential Election 2012**

---

[1] Please note that this is the preliminary draft of a paper and there may be grammatical and linguistic flaws which would be corrected subsequently. Please pardon the author for any errors or accidental omissions that might have occurred in the writing.

The US presidential election has been dominated by a variety of domestic issues since 2000. The US presidential election 2012 is strongly dominated by three main issues, the poor economic conditions, the role of US in the global stage and domestic policies by Obama. The campaign started with the incumbent having a huge edge over his competitor which was narrowed in the following months. The polls (Telegraph, 2012; Huffington Post, 2012) both illustrated very tight race in many states. The campaign is considered to be the most expensive and divisive in the past decade.

One of the key feature of this campaign is the divisive states and how they are geographically located. The race for the majority 270 electoral votes is decided on the geographical placement of the base. Obama relies mainly on the coastal states and Mid-West for support compared to Romney who relies mainly on the Southern states and prairie areas. Obama also received praises and favorable polls in international open polls across different countries (Yahoo!, 2012). Another interesting feature of the election is its heavy reliance on ideology to influence and convince the voters. The fight over ideology has been bitter and divisive. The right wing stance of the republican is strongly contrasted with the left wing nature of democrats and there are no common grounds.

The key election issue is the US economics which has pitched the two parties and candidates against one another. The weak economy has been exploited by Mitt Romney to deliver a powerful blow to the incumbent campaign during the first presidential debate. Even with new employment figure showing sign of progress, the economy is still weak and job market is poor. While the economy and job market are critical, other social issues such as immigration and social care are also important for the candidates. These issues were clearly demonstrated in the senate elections and house of representative elections. Wealth inequality has increased while unemployment remains at a high level of 9 percent with 16 million homeowners having difficulties servicing their loans (Hall, 2012). The bailout of the banks which cost billions of dollars did not make the Americans happy about their financial and job situation. This results in a negative political and electoral climate for President Obama who is blamed for the performance of the US economy.

There were other critical factors such as the race issue which has plagued the political situation in US. Recent studies uncovered interesting voter's behavior based on race preferences. These researches suggest that anti-Black attitudes did had an effect on the 2008 U.S. presidential election. Different models using various data sources and assumptions estimated two to twelve additional percentage points of the vote for President Obama if anti-Black attitudes is neutral instead (Highton, 2011; Hutchings, 2010; Jackman & Vavreck, 2011; Lewis-Beck & Tien, 2008; Piston, 2010; Schaffner, 2011). There were estimates that Mr. Obama may have lost about 5 percentage points of the popular vote in 2008 due to anti-Black attitudes (Pasek et al. 2009). Data collected suggest that anti-Black attitudes is more pervasive since the beginning of the Obama presidency. Most worryingly, these attitudes appeared at various levels among the Democrats, independents, and Republicans (Pasek et. al., 2012).

In the next section, we will discuss about the past literature on electoral result prediction.

**Literature Review**

The growth of twitter has interested researchers from various disciplines. There have been extensive of publications in this area notably in marketing as well as computer science. Some researchers study the effect of social media on the market (Honeycutt and Herring 2009; Nielsen Media Research, 2009) and found a huge variation in the intensity and usage of twitter. They also categorized the uses of twitters ranged from conversations (Honeycutt and Herring 2009) to word of mouth marketing (Jansen et al. 2009). The research literature focuses on the generic nature of twitter operating in a functional role that is not specifically specialized to evaluate political themes (Tumasjan et. Al., 2010).

There are wide spread discussions and research about the use of web forums, blogs and twitter as alternative form of political debate and information dissemination. Most researchers acknowledged the quality of the more prominent political blogs (Woodley, 2008) while others doubted the capabilities of the blogs (Sunstein, 2008). Research has also shown that while there are active participation in many of the political discussion forum (Fong, 2011) as well as blogs (Jansen and Koop 2009), the population actively participating is very small. At the same time, there was no additional information about the overall relevance of twitter in this case (Tumasjan et. Al., 2010).

Most of the current literature are focused on the effect of social media on the actual population for issues such as politics, public policies and causes. The literature covered acknowledged the lack of recognition for the non-online population influence on the political landscape (Drezner and Farrell, 2008). Several case studies have found that the online information has been quite successful acting as indicator for electoral success. (Williams and Gulati, 2008). However, there are also literature indicating that predicting electoral results using twitter is futile (Gayo-Avello, 2012). There are literatures which questioned the validity of the prediction models (Gayo-Avello, 2012; Metaxas, Mustafaraj and Gayo-Avello, 2011). The core criticisms of the various electoral predictions are listed below (Gayo-Avello, 2012),

1. Most of the researches focus on presenting the results after the election when the result is already known. This weakens the proposition that the model is predicting well.
2. The effect of incumbent is not measured in the researches.
3. No unified approach to modeling of tweets and sentiment analysis.
4. No common basis for comparison.
5. Sentiment analysis is applied as is without proper understanding of the mechanism.
6. All the tweets are assumed to be trustworthy even with the knowledge of astro-turfing behavior.
7. Demographics information is not used.
8. Political twitter data are produced only by those politically active.

In Choy et. al. (2011), a new framework incorporating socio-demographics and census information was used in conjunction with sentiment analysis to produce a model which was able to predict the Singapore Presidential Election 2011. In the following section, we will describe the application of the same framework to the US Presidential election 2012 and the result.

**Data and Methodology**

We collected 7,541,470 tweets from the start of the campaign on 12th August 2012 to 31st October 2012. Data were collected during this period of time as the Republican final nomination occurred on 11th August 2012 where the chosen Republican candidate received the endorsement by the party and nominates his vice-president. The campaign officially began on 12th August 2012 till 5 November. Even though there was a clear trend of who was going to be the Republican nominee after May, data prior to 12th August do not clearly indicate the choices available to the electorate. The data were collected through the use of the unique names of the two presidential candidates.

The amount of data collected is quite large given the length as well as the population of US. The campaign has generated almost 75 times more data as compared to the 104,003 tweets collected in the German election campaign which lasted a month (Tasmujan et. al., 2010). There were attempts to aggregate data collection in Spanish due to Obama's influence on the Hispanic population. However, the authors are not proficient in Spanish and could not translate the tweets to English.

This resulted in the decision to use English tweets as the only source of data. Proper data cleaning were done to ensure that a proper and unadulterated collection of tweets with proper chronological order is available for analysis.

To extract the sentiments from the data automatically, a generic purpose corpus was used for this analysis. While there were several corpus and programs available online to conduct sentiment analysis, AFINN (Nielsen, 2011) was considered by the authors to be the most suitable one for analyzing within this context and can be applied easily to any textual data to calculate the sentiment values. It is important to note that this corpus was not modified from the original version presented by the original author for the purpose of analyzing this campaign in order to ensure that the model can be replicated in other elections. Socio-demographics data and census information (US Census, 2012) were collected to correct the bias in the online data. We have assumed that people who tweet were truthful in expressing their views to avoid further complications in the estimation of the sentiment as well as the choice of candidate. Below is a time series chart showing the positive and negative sentiment for both candidates during the period of time.

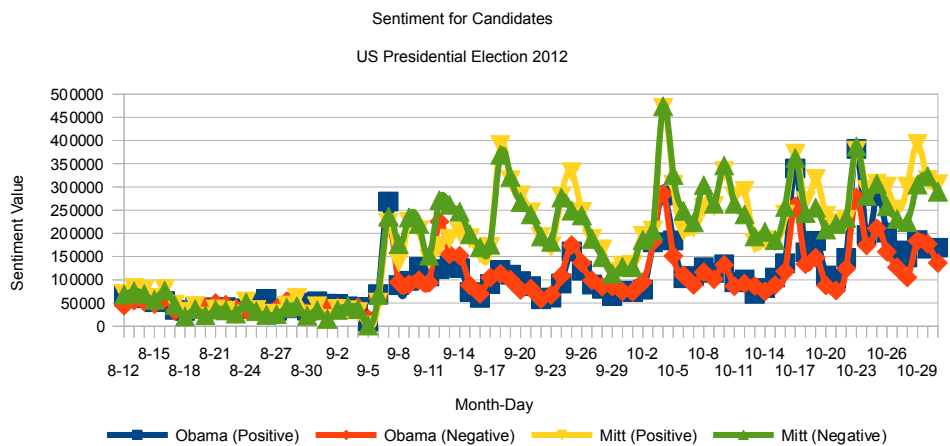

The original model described in Choy et al. (2011) uses a combination of age data, computer literacy as well as prior election information to calculate the vote share. However, due to the amount of information that is available to the authors, the model has to be modified slightly to use the information. The model assumes the following electorate structure which is specific to each election due to the nature of each post and country.

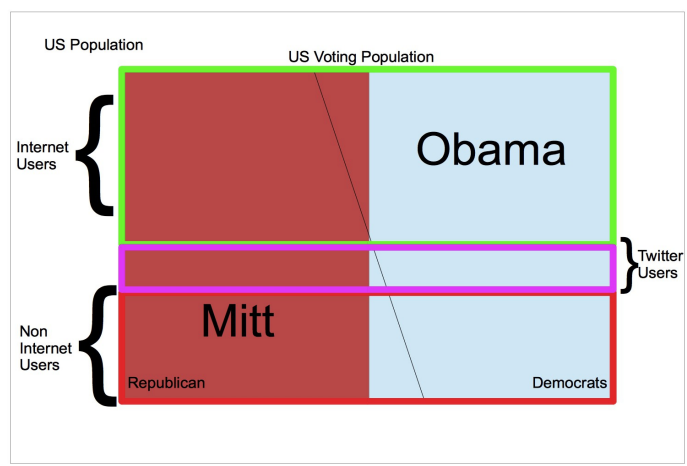

The model separates the population into several components. Below we will define the various components in the population

Let i = index representing the states for i = 1, ..., 52 (52 states of America)
Let j = index of representing the candidates for j = 1, 2 (1 = Obama, 2 = Mitt)
Let k = index representing the party for k = 1, 2 (1 = Democrats, 2 = Republicans)

However, in the case of the US Presidential Elections, age level information is not available. There are two additional pieces of information that are available for use. The first piece of information is the percentage of the US population using twitter. The second piece of information is the detailed state by state break down of the internet usage (US Census, 2012). These two pieces of information provide additional insights to the influence of social media on the population on a more granular level.

Let $C_i$ = percentage of people in state i who uses internet but not twitter (52 states of America)
Let $S_i$ = percentage of people in state i who uses twitter (52 states of America)

In the case of the US electoral climate, we have pretty detailed information about the prior elections information and declared party affiliations for the voters.

Let $P_{ik}$ = percentage of people in state i who supported party k
Let $E_j$ = percentage of people from twitter who supported candidate j

This two information provides us with two different models. The first model assumes that the twitter information reflects the general online netizens' belief and the sentiment from twitter can be used to calculate the support for the candidates.

Support for Candidate j for state i for twitter user = $S_i E_j$
Support for Candidate j for state i for internet but not twitter user = $C_i E_j$
Support for Candidate j for state i for non internet user = $(1-C_i)P_{ik}$

$$\text{Total Support for Candidate } j = \sum_{i=1}^{52} \left( S_i E_j + C_i E_j + (1-C_i) P_{ik} \right) \qquad (1)$$

The second model assumes that the twitter influence is limited to twitter population and that the rest of the electorate are better modeled by their prior political affiliation.

Support for Candidate j for state i for twitter user = $S_i E_j$
Support for Candidate j for state i for internet but not twitter user = $C_i P_{ik}$
Support for Candidate j for state i for non internet user = $(1-C_i)P_{ik}$

$$\text{Total Support for Candidate } j = \sum_{i=1}^{52} \left( S_i E_j + C_i P_{ik} + (1-C_i) P_{ik} \right) \qquad (2)$$

Comparison between the two models will determine whether twitter has effect on the rest of the internet and measure the amount of influence twitter has on the general online population. This will address some of the questions raised by Gayo-Avello (2011).

**Result**

In the first model, we will assume that the twitter sentiment reflects the overall online population sentiment. Using the information, below are the results.

| State | Obama | Mitt | State | Obama | Mitt | State | Obama | Mitt |
|---|---|---|---|---|---|---|---|---|
| Alabama | 47.38% | 52.62% | Louisiana | 47.77% | 52.23% | Oklahoma | 46.75% | 53.25% |
| Alaska | 48.97% | 51.03% | Maine | 51.73% | 48.27% | Oregon | 51.27% | 48.73% |
| Arizona | 49.52% | 50.48% | Maryland | 52.32% | 47.68% | Pennsylvania | 51.29% | 48.71% |
| Arkansas | 47.03% | 52.97% | Massachusetts | 52.17% | 47.83% | Rhode Island | 52.90% | 47.10% |
| California | 52.07% | 47.93% | Michigan | 51.74% | 48.26% | South Carolina | 48.98% | 51.02% |
| Colorado | 50.96% | 49.04% | Minnesota | 51.00% | 49.00% | South Dakota | 49.32% | 50.68% |
| Connecticut | 52.23% | 47.77% | Mississippi | 48.28% | 51.72% | Tennessee | 48.01% | 51.99% |
| Delaware | 52.81% | 47.19% | Missouri | 50.15% | 49.85% | Texas | 49.06% | 50.94% |
| District of Columbia | 58.40% | 41.60% | Montana | 49.63% | 50.37% | Utah | 48.81% | 51.19% |
| Florida | 50.52% | 49.48% | Nebraska | 48.85% | 51.15% | Vermont | 53.20% | 46.80% |
| Georgia | 49.71% | 50.29% | Nevada | 51.14% | 48.86% | Virginia | 50.84% | 49.16% |
| Hawaii | 54.90% | 45.10% | New Hampshire | 50.90% | 49.10% | Washington | 51.23% | 48.77% |
| Idaho | 48.12% | 51.88% | New Jersey | 51.57% | 48.43% | West Virginia | 48.27% | 51.73% |
| Illinois | 52.71% | 47.29% | New Mexico | 51.90% | 48.10% | Wisconsin | 51.37% | 48.63% |
| Indiana | 50.28% | 49.72% | New York | 52.84% | 47.16% | Wyoming | 47.59% | 52.41% |
| Iowa | 51.12% | 48.88% | North Carolina | 50.26% | 49.74% | Popular Vote | 50.71% | 49.29% |
| Kansas | 49.06% | 50.94% | North Dakota | 49.23% | 50.77% | Electoral Vote | 67.35% | 32.65% |
| Kentucky | 47.81% | 52.19% | Ohio | 50.63% | 49.37% | | | |

From the result, we can see that the race is pretty tight in most states. The swing states such as Colorado, Florida, Iowa and Ohio have margin smaller than 1%. We can see that Indiana is also poised to be Democrat than Republican which is contrary to the polls. This can be due to the overwhelming weighting given to internet population which might not be entirely appropriate. However, the final result for Obama stands at 50.1% which is adjusted to 50.75% for two party scenario and the absolute error from the model is 0.04%. This is interesting and perplexing as the model did not perform very well for the state level model.

Let us observe the second model which models the online population based on prior party affiliation.

| State | Obama | Mitt | State | Obama | Mitt | State | Obama | Mitt |
|---|---|---|---|---|---|---|---|---|
| Alabama | 39.91% | 60.09% | Louisiana | 40.98% | 59.02% | Oklahoma | 35.96% | 64.04% |
| Alaska | 39.14% | 60.86% | Maine | 56.97% | 43.03% | Oregon | 56.11% | 43.89% |
| Arizona | 45.64% | 54.36% | Maryland | 60.77% | 39.23% | Pennsylvania | 54.08% | 45.92% |
| Arkansas | 40.02% | 59.98% | Massachusetts | 60.27% | 39.73% | Rhode Island | 61.62% | 38.38% |
| California | 59.95% | 40.05% | Michigan | 56.73% | 43.27% | South Carolina | 45.45% | 54.55% |
| Colorado | 53.33% | 46.67% | Minnesota | 53.69% | 46.31% | South Dakota | 45.31% | 54.69% |
| Connecticut | 59.57% | 40.43% | Mississippi | 43.74% | 56.26% | Tennessee | 42.68% | 57.32% |
| Delaware | 60.79% | 39.21% | Missouri | 49.40% | 50.60% | Texas | 44.35% | 55.65% |
| District of Columbia | 88.25% | 11.75% | Montana | 47.58% | 52.42% | Utah | 36.00% | 64.00% |
| Florida | 50.97% | 49.03% | Nebraska | 42.48% | 57.52% | Vermont | 65.75% | 34.25% |
| Georgia | 47.33% | 52.67% | Nevada | 54.67% | 45.33% | Virginia | 52.41% | 47.59% |
| Hawaii | 69.35% | 30.65% | New Hampshire | 53.75% | 46.25% | Washington | 56.93% | 43.07% |
| Idaho | 37.52% | 62.48% | New Jersey | 56.58% | 43.42% | West Virginia | 43.37% | 56.63% |
| Illinois | 60.77% | 39.23% | New Mexico | 56.26% | 43.74% | Wisconsin | 55.63% | 44.37% |
| Indiana | 49.99% | 50.01% | New York | 61.04% | 38.96% | Wyoming | 34.30% | 65.70% |
| Iowa | 53.57% | 46.43% | North Carolina | 49.90% | 50.10% | Popular Vote | 52.47% | 47.53% |
| Kansas | 42.53% | 57.47% | North Dakota | 45.20% | 54.80% | Electoral Vote | 60.82% | 39.18% |
| Kentucky | 42.09% | 57.91% | Ohio | 51.39% | 48.61% | | | |

The second model predicts a comfortable 52.47% win for Obama. Several of the results for state level information is fairly close to the open polls on the internet provided by various agencies (Huffington Post, 2012; Telegraph, 2012) prior to the election. The results were compared to the actual election results as well as the base line values computed by Daniel Gayo-Avello using absolute error computed.

| State | Predicted | Actual | AE (Actual) | Baseline | AE (Baseline) |
|---|---|---|---|---|---|
| Alabama | 39.91% | 38.74% | 1.17% | 39.11% | 0.37% |
| Alaska | 39.14% | 42.81% | 3.67% | 38.94% | 3.87% |
| Arizona | 45.64% | 43.85% | 1.79% | 45.69% | 1.84% |
| Arkansas | 40.02% | 37.87% | 2.15% | 39.83% | 1.96% |
| California | 59.95% | 58.40% | 1.55% | 62.28% | 3.88% |
| Colorado | 53.33% | 51.63% | 1.70% | 54.55% | 2.92% |
| Connecticut | 59.57% | 57.89% | 1.68% | 61.32% | 3.43% |
| Delaware | 60.79% | 59.44% | 1.35% | 62.64% | 3.20% |
| District of Columbia | 88.25% | 92.77% | 4.52% | 93.40% | 0.63% |
| Florida | 50.97% | 50.40% | 0.57% | 51.42% | 1.02% |
| Georgia | 47.33% | 45.94% | 1.39% | 47.37% | 1.43% |
| Hawaii | 69.35% | 72.84% | 3.49% | 72.99% | 0.15% |
| Idaho | 37.52% | 31.37% | 6.15% | 36.98% | 5.61% |
| Illinois | 60.77% | 58.24% | 2.53% | 62.73% | 4.49% |
| Indiana | 49.99% | 44.64% | 5.35% | 50.52% | 5.88% |
| Iowa | 53.57% | 52.92% | 0.65% | 54.85% | 1.93% |
| Kansas | 42.53% | 39.04% | 3.49% | 42.39% | 3.35% |
| Kentucky | 42.09% | 38.45% | 3.64% | 41.77% | 3.32% |
| Louisiana | 40.98% | 40.49% | 0.49% | 40.54% | 0.05% |
| Maine | 56.97% | 58.32% | 1.35% | 58.83% | 0.51% |
| Maryland | 60.77% | 62.34% | 1.57% | 62.93% | 0.59% |
| Massachusetts | 60.27% | 61.94% | 1.67% | 63.20% | 1.26% |
| Michigan | 56.73% | 53.75% | 2.98% | 58.37% | 4.62% |
| Minnesota | 53.69% | 53.14% | 0.55% | 55.23% | 2.09% |
| Mississippi | 43.74% | 44.04% | 0.30% | 43.36% | 0.68% |
| Missouri | 49.40% | 45.13% | 4.27% | 49.93% | 4.80% |
| Montana | 47.58% | 44.36% | 3.22% | 48.83% | 4.47% |
| Nebraska | 42.48% | 38.34% | 4.14% | 42.39% | 4.05% |
| Nevada | 54.67% | 53.49% | 1.18% | 56.39% | 2.90% |
| New Hampshire | 53.75% | 52.66% | 1.09% | 54.87% | 2.21% |
| New Jersey | 56.58% | 58.79% | 2.21% | 57.86% | 0.93% |
| New Mexico | 56.26% | 55.21% | 1.05% | 57.66% | 2.45% |
| New York | 61.04% | 63.14% | 2.10% | 63.58% | 0.44% |
| North Carolina | 49.90% | 48.91% | 0.99% | 50.17% | 1.26% |
| North Dakota | 45.20% | 40.04% | 5.16% | 45.59% | 5.55% |
| Ohio | 51.39% | 50.96% | 0.43% | 52.33% | 1.37% |
| Oklahoma | 35.96% | 33.16% | 2.80% | 34.35% | 1.19% |
| Oregon | 56.11% | 55.01% | 1.10% | 58.41% | 3.40% |
| Pennsylvania | 54.08% | 52.59% | 1.49% | 55.23% | 2.64% |
| Rhode Island | 61.62% | 64.21% | 2.59% | 64.20% | 0.01% |
| South Carolina | 45.45% | 43.43% | 2.02% | 45.46% | 2.03% |
| South Dakota | 45.31% | 40.60% | 4.71% | 45.70% | 5.10% |
| Tennessee | 42.68% | 39.62% | 3.06% | 42.37% | 2.75% |
| Texas | 44.35% | 41.91% | 2.44% | 44.06% | 2.15% |
| Utah | 36.00% | 25.53% | 10.47% | 35.47% | 9.94% |
| Vermont | 65.75% | 68.30% | 2.55% | 68.90% | 0.60% |
| Virginia | 52.41% | 51.17% | 1.24% | 53.18% | 2.01% |
| Washington | 56.93% | 56.35% | 0.58% | 58.75% | 2.40% |
| West Virginia | 43.37% | 36.38% | 6.99% | 43.33% | 6.95% |
| Wisconsin | 55.63% | 52.50% | 3.13% | 57.06% | 4.56% |
| Wyoming | 34.30% | 28.62% | 5.68% | 33.44% | 4.82% |
| MAE | | | 2.60% | | 2.75% |

From the comparison, we can see that the MAE of model 2 is lower than the baseline model. At the same time, from the comparison, we can see that the AEs are lower for the key swing states for model 2 as compared to the baseline model. However, the final popular vote result yields 52.4% for Obama which is much higher than model 1 and a higher absolute error of 1.65% which is undesirable.

**Conclusion**

From the results, both models indicated that the incumbent, President Obama should win the 2012 Presidential election. This prediction is validated by the final results released by various agencies. Forecasted value for the final popular vote percentage for Obama is fairly accurate for model 1 which models the internet using population using the sentiment collected from the twitter. However, the state level modeling is better served by model 2. This raised questions about the validity of framework and some of the problems that were raised by other authors (Gayo-Avello, 2012; Metaxas, Mustafaraj and Gayo-Avello, 2011). Below are some of the questions derived from the papers as well as reviewers.

*The effect of incumbent is not measured in the researches.*

To be able to measure the effect of incumbent, the pre-requisite is to have an incumbent. In a previous discussion with a reviewer, the recommendation for the case of no incumbent is to use party association as a measure. In the case of US Presidential Elections 2012, there is an incumbent and it is possible to measure the effect of incumbent. Comparing model 1 and model 2, we can observe that the state by state level prediction is far better in model 2 which indicates that the prior election orientation is retained and does have an impact to the accuracy of the forecast. However, this effect might be only applicable to Presidential elections and as this is not tested in other elections.

*No common basis for comparison.*

To have a common basis for comparison, there needs to be a baseline model. In some of the papers mentioned, the recommended approach is to compare the final result with the prior election results as a baseline. If the previous election can be used to predict the result better than the model, than the model has performed poorly (Gayo-Arvello, 2012). This is particularly important given the previous conclusion that there is an incumbent effect.

*All the tweets are assumed to be trustworthy even with the knowledge of astro-turfing behavior.*

The author is of the opinion that we have to assume that the tweets are trustworthy as it will be too much work to attempt verifying the veracity of each tweet. As for the astro-turfing behavior, it will be prohibitively costly to attempt cleaning all the information out through human labor. There have been various attempts to model astro-turfing activities through statistical modeling with varying degree of success. However, these attempts still requires training samples which is constructed through human labour. In this case, the author argues that the effect of astro-turfing for any political elections will be neutralized in the cases where both parties are equally matched in terms of resources. For the case of US presidential elections, both candidates have substantial amount of resources available for them to attempt serious level of astro-turfing activities. If any one candidate attempts to flood the social media platform with positive sentiments for his camp, it is likely to address with the same magnitude by the other candidate. Thus in this scenario, given the similar resources shared, it is likely that astro-turfing activities from both parties will cancel each other.

*Demographics information is not used.*

In this case, demographics information has been implemented in a different manner as compared to the one described in the literature. Demographics information are very important as they are factors which primarily drive the voting behavior. One important demographics information is the use of race and gender. In the case of the US presidential elections 2012, the blacks and hispanics voted for Obama overwhelmingly. However, these information are not always available at levels which are sufficiently granular for modeling purposes. In this case, we used the internet and twitter distribution information on a state level to indicate the spread of influence for twitter sentiments. However, one of the key difference from prior model (Choy et. al., 2011) is the lack of the age level information which helps to differentiate the internet and twitter usage. While there are useful information about the percentage of support for each party for each age group for the entire nation, there was such information at the state level making it difficult to model the effects of the demographics. From the results of the US presidential elections 2012, we demonstrated the importance of the demographics information in predicting the results on the state level where the MAE is very well controlled. This addressed the concerns raised by various researchers on the importance of demographics information to the accuracy of the results (Gayo-Avello, 2012; Metaxas, Mustafaraj and Gayo-Avello, 2011). The combination of demographics level information with twitter in model 2 produced results which were better than the baseline level by 5%. While the result was not an extreme improvement, it does point to the direction that better methodology developed incorporating both data will yield better results.

However, model 1 produced results which are closer to the final tally for the election. The authors were very surprised at the accuracy of the result with the final tally standing at 50.6% (51.4% converted to the two party tally) which differs from the predicted result at 50.71% by 0.11% ( 0.7%) . The accuracy of the model 1 over model 2 which is a far more accurate model on the overall state by state level information indicated several interesting insights. The twitter sentiments does in a way affect internet users' perception even if they are not users' of twitter. The twitter sentiment also acts as good barometer of the electorate's opinion of the candidates. Given the increasing number of users embracing twitter, the future electorate will be more likely to express opinions online through twitter and other social media channels making such measurements likely to be more accurate than before.

*Political twitter data are produced only by those politically active.*

In the US presidential elections, voting is not mandatory for the electorate and thus only the politically active or motivated people will be voting. In this case, political twitter data will be relevant and most likely a product of the electorate who will be voting. The author acknowledged that this scenario might not have occurred in other elections. However, prior research (Choy et. al., 2011) has indicated its relevance even in mandatory voting environment. This is partially linked to the previous factor of twitter influence on the internet using population.

*The political tweets were not collected on a geographical basis.*

The authors acknowledged that the tweets were not collected on a geographical basis and thus might have included information or tweets from other regions. However, the authors argued that most of the tweets were not tagged geographically and thus is almost impossible to have collected a representative sample of tweets from twitter without collecting the untagged ones. At the same time, even though an election such as the US presidential election is monitored globally, most of the tweets will originate from US and not other regions as they are not the key actors in the elections.

Thus, from the results of the model, we can see that if there are any such effects, the effects will be minimum and does not affect the prediction.

The results were released prior to the final results and uploaded to arXiv for verification as well as notifying 5 independent observers who are not participants of the elections for the purpose of establishing this paper as a prediction rather than a post event analysis. Several questions raised have been answered by the models. The model has performed well using very limited data available to the author and the absolute error is lower than the baseline models. This paper has again validated the possibilities of using twitter information to predict elections. Future research directions will be to build an adaptive model to filter astro-turfing as well as calculation of the age level party supports to improve the prediction accuracy.

## Acknowledgement

The authors will like to thank the various independent observers for assisting to validate the results. The authors will also wish to acknowledge the advices and opinions rendered by Daniel Gayo-Arvello with regards to the model presented in this paper.